\documentclass[11pt]{article}
\usepackage{hyperref}
\pdfoutput=1
\begin{document}
\title{The blistering of a viscoelastic filament\\ of a droplet of saliva}
\author{Christian Wagner$^1$, Rainer Sattler$^1$, and $^2$Jens Eggers\\
\\\vspace{6pt} $^1$Technical Physics, Saarland University, 66123 Saarbruecken, Germany\\
\\\vspace{6pt} $^2$School of Mathematics, University of Bristol,\\ University
Walk, Bristol BS8 1TW United Kingdom
}
\maketitle
\begin{abstract}
A fluid dynamics video of the break up of a droplet of saliva is shown. First a viscoelastic filament is formed and than the blistering of this filament is shown. Finally, a flow induced phase separation takes place nanometer sized solid fiber remains that consist out of the biopolymers.
\end{abstract}
\section{Introduction}


The \href{http://hdl.handle.net/1813/13680}{video} shows the shadowgraph of a break-up of a capillary bridge of a droplet of saliva from a healthy donor. The droplet is placed between two steel plates that are gently drawn apart until the capillary bridge becomes unstable. The videos are rotated 90$^\circ$ to the right. In the first part, an overview is given of the formation of the elastic filament that blisters finally into a series of beads. The formation of the filament is related to the high elongational viscosity of saliva that originates from the polymeric biopolymers [1]. Second, it is shown that on the highly stretched filament a Rayleigh-Plateau like instability can grow [2]. The third part shows the final rupture of the filament. Apparently, the thin filaments that connect the remaining beads have a very high bending modulus. If they are captured on a metallic carrier it is possible tho show that they consist out of polymers only, which is shown in the fourth part of the movie.

\begin{enumerate}

\item E. Zussman, A.L. Yarin, R.M. Nagler, J. Dental Res. \textbf{86}, 281 (2007)
\item   R. Sattler, C. Wagner, J. Eggers, Phys. Rev. Lett. \textbf{100}, 164502 (2008).
\end{enumerate}

\end{document}